\documentclass[11pt,twoside]{article}
\usepackage{asp2014}

\aspSuppressVolSlug
\resetcounters
\begin{document}

\title{The AGILEScience App to execute gamma-ray scientific analyses from mobile devices}

\author{A.~Bulgarelli,$^1$ N.~Parmiggiani,$^1$ V.~Fioretti,$^1$, L.~Baroncelli,$^1$ A.~Addis,$^1$ A.~Di Piano,$^1$  C.~Pittori,$^{2,3}$ and M.~Tavani,$^{2}$}
\affil{$^1$INAF OAS Bologna, Via P. Gobetti 93/3, 40129 Bologna, Italy. \email{andrea.bulgarelli@inaf.it}}

\affil{$^2$INAF/OAR Roma, Via Frascati 33, I-00078 Monte Porzio Catone, Roma, Italy}

\affil{$^3$ASI/SSDC Roma, Via del Politecnico snc, I-00133 Roma, Italy.}


\paperauthor{Andrea~Bulgarelli}{andrea.bulgarelli@inaf.it}{0000-0001-6347-0649}{INAF}{OAS}{Bologna}{BO}{40129}{Italy}
\paperauthor{Nicol\`{o}~Parmiggiani}{nicolo.parmiggiani@inaf.it}{0000-0002-4535-5329}{INAF}{OAS}{Bologna}{BO}{40129}{Italy}
\paperauthor{Valentina~Fioretti}{valentina.fioretti@inaf.it}{0000-0002-6082-5384}{INAF}{OAS}{Bologna}{BO}{40129}{Italy}
\paperauthor{Leonardo~Baroncelli}{leonardo.baroncelli@inaf.it}{0000-0002-9215-4992}{INAF}{OAS}{Bologna}{BO}{40129}{Italy}
\paperauthor{Antonio~Addis}{antonio.addis@inaf.it}{0000-0002-0886-8045}{INAF}{OAS}{Bologna}{BO}{40129}{Italy}
\paperauthor{Marco~Tavani}{marco.tavani@inaf.it}{0000-0003-2893-1459}{INAF}{IAPS}{Roma}{RO}{00133}{Italy}




\begin{abstract}
AGILE is a space mission launched in 2007 devoted to high-energy astrophysics. The AGILE Team is involved in the multi-messenger campaigns to send and receive science alerts about transient events in the shortest time possible. For this reason, the AGILE Team developed several real-time analysis pipelines to analyse data and follow-up external science alerts. However, the results obtained by these pipelines are preliminary and must be validated with manual analyses that are the bottleneck of the workflow. To speed up the scientific analysis performed by scientists, the AGILE Team developed the AGILEScience mobile application (for iOS and Android devices) that offers to the AGILE Team a password-protected section used to visualise the results of automated pipelines. We present in this contribution an improved functionality of the AGILEScience application that aims to enable the AGILE Team to execute a full scientific analysis using their mobile devices. When the analysis is completed, the system sends an email to notify the user that can visualise the results (e.g. plots, tables, and HTML pages) through the application. The possibility to perform scientific analysis from a mobile device enables the AGILE researchers to perform fast scientific analyses remotely to validate the preliminary results obtained with the automated pipelines. This workflow reduces the overall reaction time of the AGILE Team for the follow-up of transient phenomena. 
\end{abstract}

\section{Introduction}

AGILE (Astrorivelatore Gamma ad Immagini LEggero - Light Imager for Gamma-Ray Astrophysics) is a space mission of the Italian Space Agency (ASI) designed for X-ray and gamma-ray observations and was launched on 23rd Apr 2007 \citep{2008NIMPA.588...52T, 2009A&A...502..995T}. The AGILE payload consists of the Silicon Tracker (ST), the SuperAGILE X-ray detector, the CsI(Tl) Mini-Calorimeter (MCAL), and an AntiCoincidence System (ACS). The combination of ST, MCAL, and ACS form the Gamma-Ray Imaging Detector (GRID). 

The AGILE Team participates in multi-messenger astronomy to follow up science alerts about transient events such as Gamma-ray Bursts (GRB) or flare from Active Galactic Nuclei (AGN). To promptly send communications about transient phenomena in the shortest time possible, the AGILE team developed several real-time analysis (RTA) pipelines \citep{2019ExA....48..199B, Parmiggiani:20214o} that execute automated scientific analyses. In addition, the AGILE RTA system is designed to react and follow up external science alerts sent by other facilities.

Usually, the results produced by the RTA system must be validated by the AGILE researchers with manual analyses before sending science alerts to the community. This task is the bottleneck of the whole workflow because to perform manual analyses, the researchers must access the AGILE data and computational servers through their PC. Moreover, the science alerts can be received at each time during the day or night, and researchers may not have easy access to the computational resources. Therefore, the first step to improve the reaction time of the AGILE Team is to make available the results of automated analyses through a password-protected web Graphical User Interface (GUI). In addition, the AGILE Team developed the AGILEScience mobile application described in the next section that can be used to visualise the results of the RTA system.

\section{The AGILEScience App}

\begin{figure}[!htb]
   \begin{minipage}{0.48\textwidth}
     \centering
     \includegraphics[width=.8\linewidth]{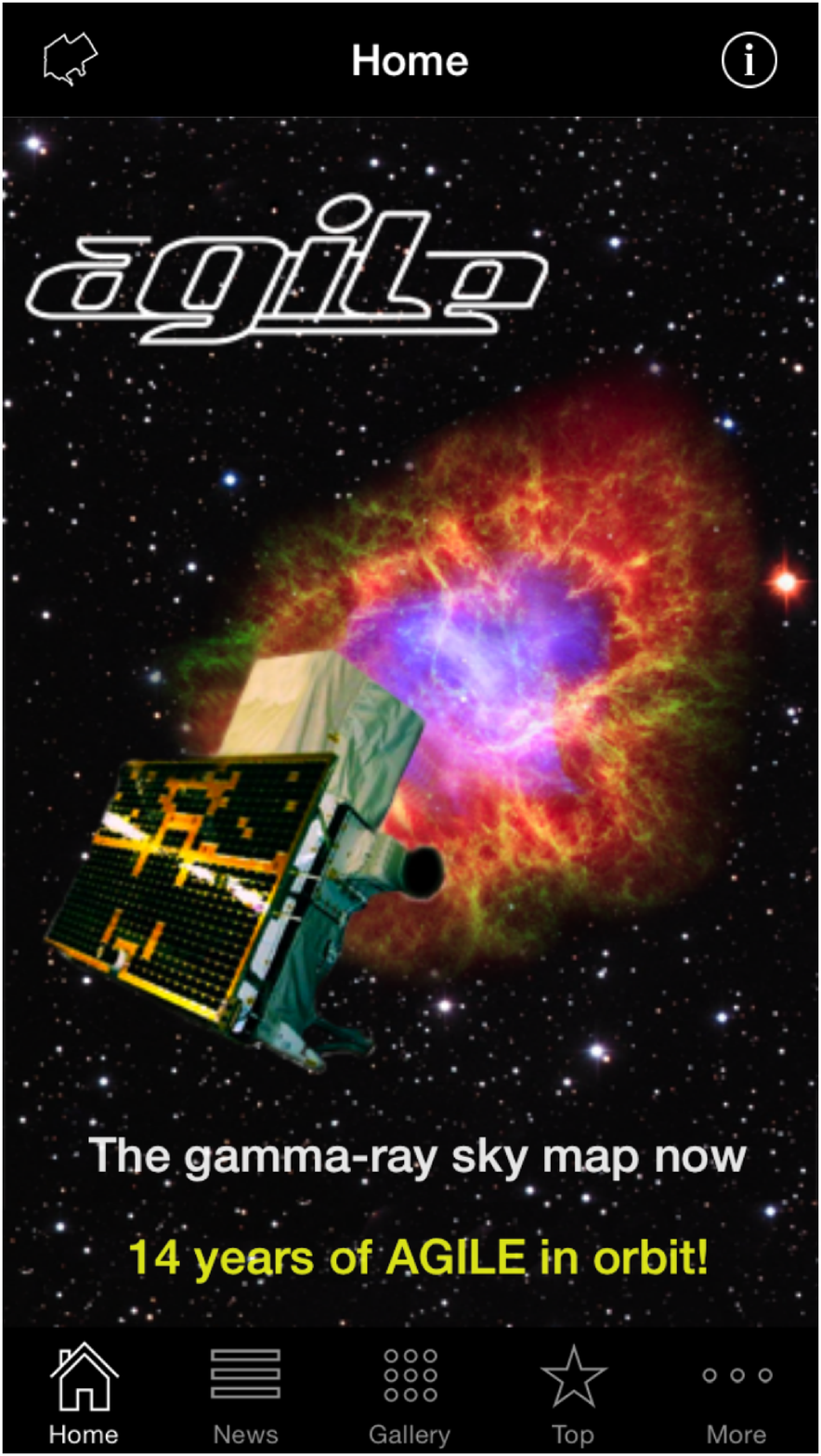}
     \caption{Home page of the App.}\label{fig:app1}
   \end{minipage}\hfill
   \begin{minipage}{0.48\textwidth}
     \centering
     \includegraphics[width=.8\linewidth]{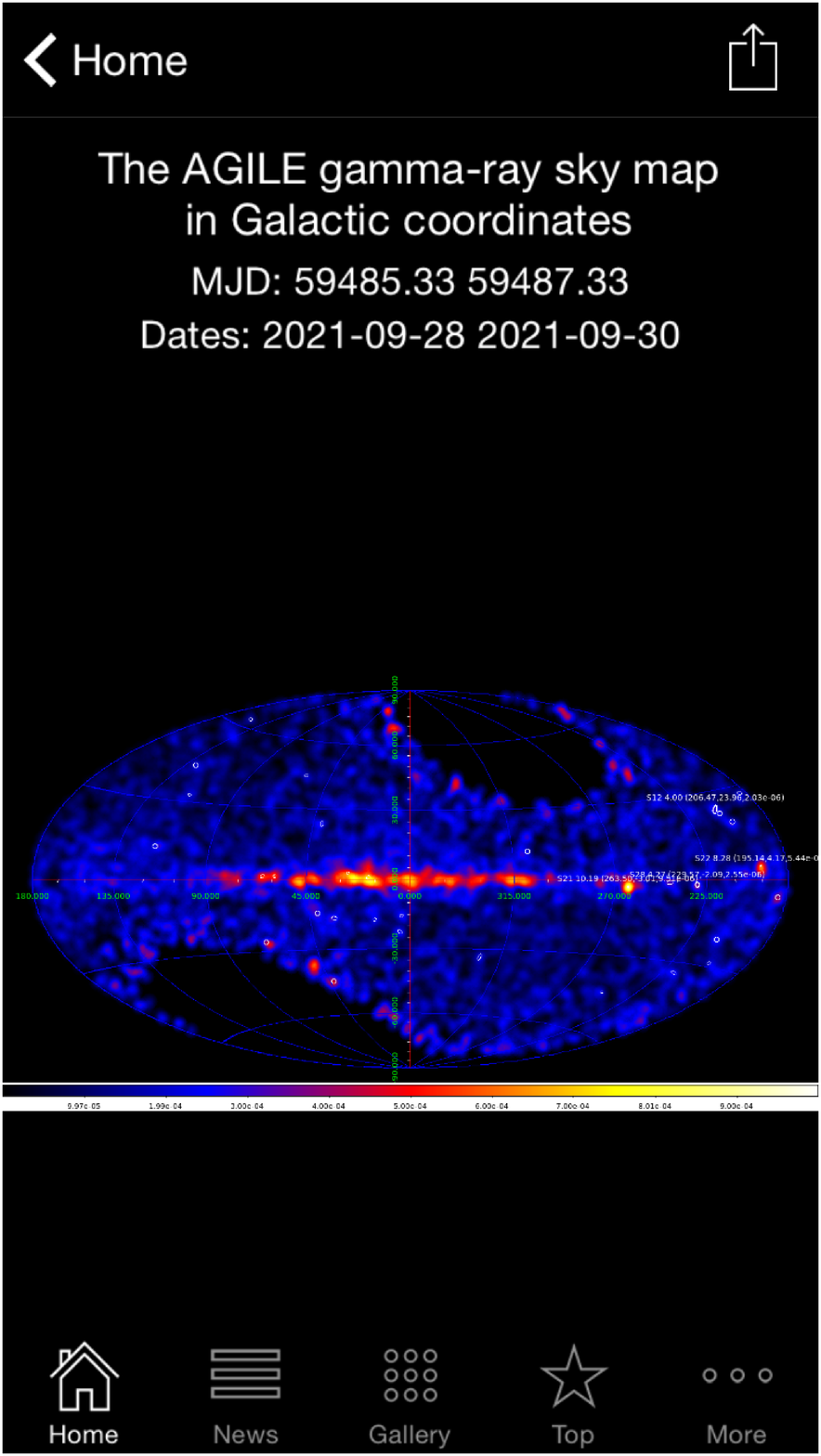}
     \caption{Gamma-ray sky map from the last data acquired by the satellite.}\label{fig:app2}
   \end{minipage}
\end{figure}

The AGILEScience application (Figure \ref{fig:app1}) is a mobile software developed for the iOS and Android Operating Systems (representing almost 100$\%$ of the mobile market), available for free from the Apple App Store for iPhone\footnote{https://apps.apple.com/app/agilescience/id587328264} and iPad\footnote{https://apps.apple.com/ao/app/agilescience-for-ipad/id690462286} and from the Google Play Store\footnote{https://play.google.com/store/apps/details?id=com.agile.science\&hl=en\_US\&gl=US}. The App has two main goals: 

\begin{enumerate}
    \item present to the public the most important AGILE scientific results, the gamma-ray sky maps updated every few hours (Figure \ref{fig:app2}), a gallery of images and videos about the construction and launch of the satellite, and other interesting information;
    \item offers to the AGILE Team a password-protected section to monitor the data flow and visualise the scientific results of the RTA pipelines. 
\end{enumerate}

The App is interfaced with a Wordpress (\url{https://wordpress.com}) website to obtain updated information for the public (e.g. the list of papers related to AGILE) and with the AGILE remote servers to retrieve the monitoring data and the analyses results (e.g. tables, plots, sky maps, etc). In addition, the WordPress website automatically updates its content, retrieving the news from the Media INAF (\url{https://www.media.inaf.it}) online newsletters and an SSDC web page.

\section{Execution of scientific analysis from the App}

We present in this contribution an improved functionality of the App that the AGILE Team can use to execute scientific analyses on GRID data directly from their mobile device, improving the reaction time during the follow-up of transient events. This new feature is implemented inside the password-protected section of the App that can be accessed only by the AGILE Team.

\begin{figure}[!htb]
   \begin{minipage}{0.48\textwidth}
     \centering
     \includegraphics[width=.8\linewidth]{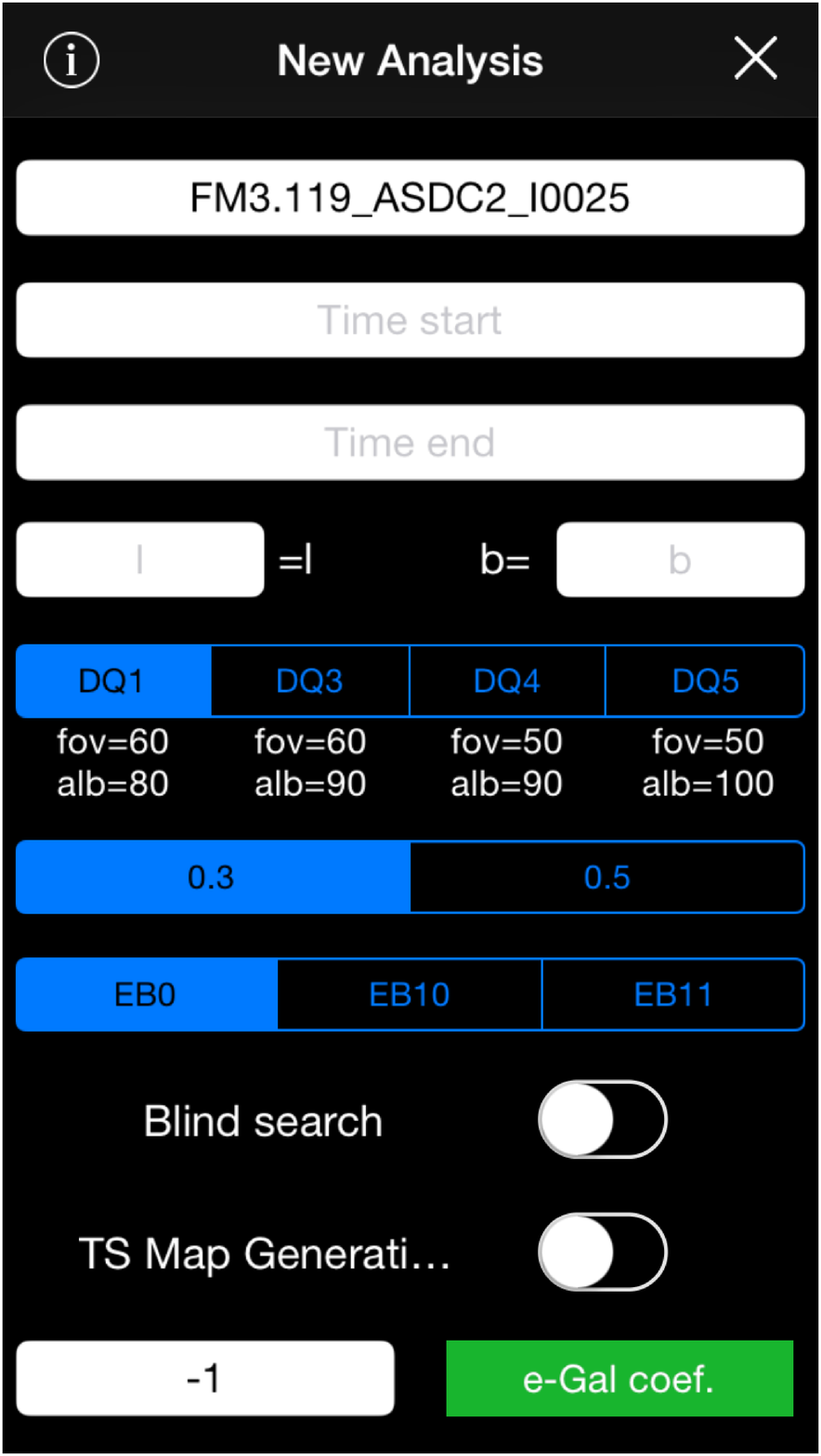}
     \caption{Predefined form to insert configuration parameters used for the analysis.}\label{fig:app3}
   \end{minipage}\hfill
   \begin{minipage}{0.48\textwidth}
     \centering
     \includegraphics[width=.8\linewidth]{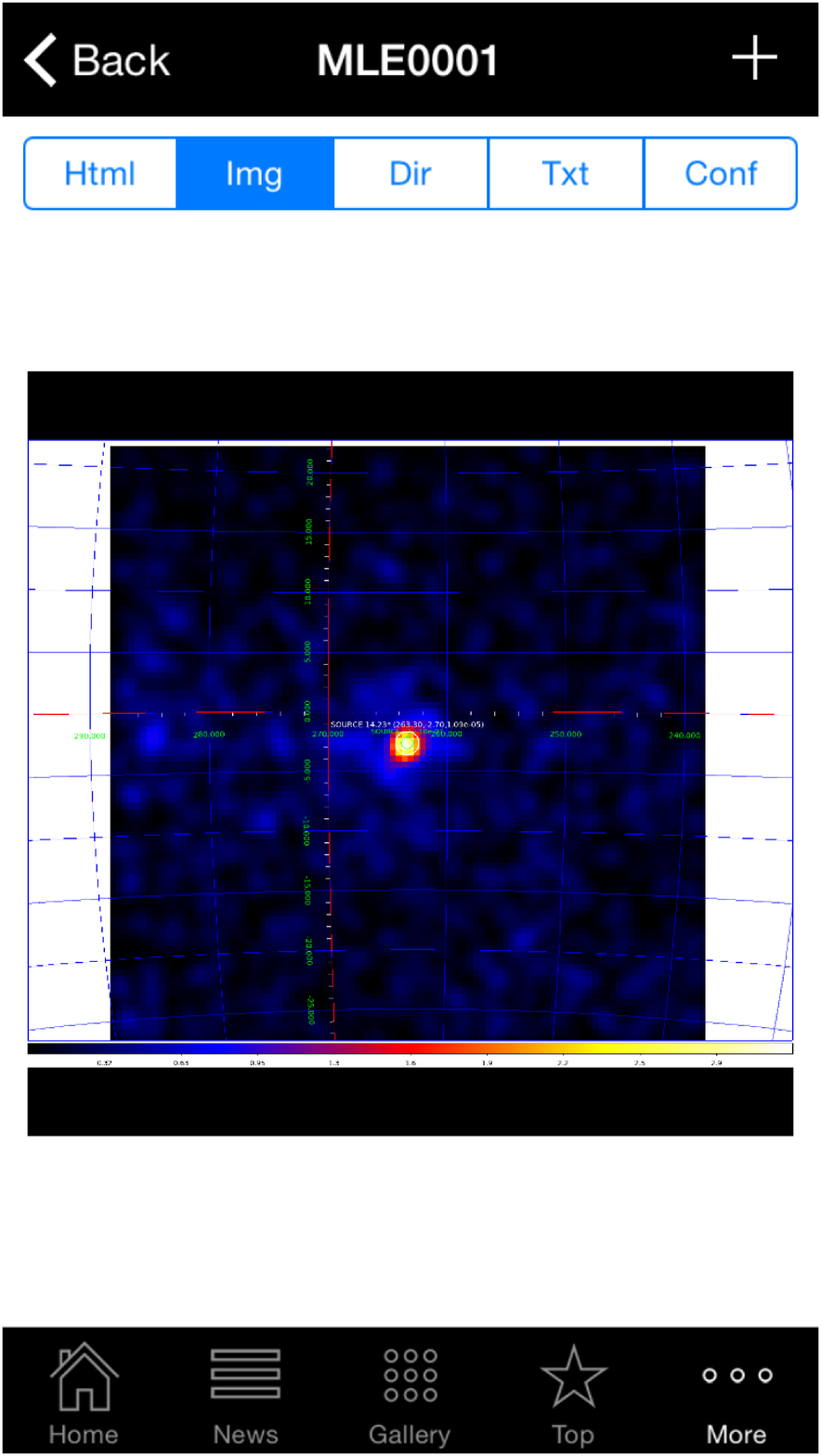}
     \caption{Example of results (a counts map) that can be visualised through the App.}\label{fig:app4}
   \end{minipage}
\end{figure}

The researchers can start a new analysis editing a predefined form to configure all the required parameters  (Figure \ref{fig:app3}). The App send the parameters configuration to the computing server that automatically executes the analysis. When the analysis is completed, the system informs the user by sending an email. The researchers can visualise the results (e.g. maximum likelihood estimator results, sky maps, etc) from the App (Figure \ref{fig:app4}). 

This workflow reduces the time required for the manual validation of scientific results to publish a science alert or follow up a transient event sent by other facilities.

\section{Conclusion}

The AGILE Team developed several RTA pipelines to execute fast analysis on the data acquired by the AGILE detectors used to follow-up transient phenomena detected by AGILE or shared by other facilities. However, the some results obtained by the RTA system must be validated by the AGILE Team with manual analysis. This procedure delays the reaction time during the follow-up of transient events.

To reduce the time required for the manual validation of the scientific results, we improved the AGILEscience App for mobile devices implementing a new functionality inside the password-protected section that the AGILE Team can use to submit analysis on AGILE/GRID data on the computing servers. The analyses results are shown directly inside the App. This workflow reduces the overall time required to validate automated results, improving the AGILE Team capability to follow up transient phenomena.

\acknowledgements The AGILE Mission is funded by the Italian Space Agency (ASI) with scientific and programmatic participation by the Italian National Institute for Astrophysics (INAF) and the Italian National Institute for Nuclear Physics (INFN). The investigation is supported by the ASI grant  I/028/12/6. We thank the ASI management for unfailing support during AGILE operations. We acknowledge the effort of ASI and industry personnel in operating the  ASI ground station in Malindi (Kenya), and the data processing done at the ASI/SSDC in Rome: the success of AGILE scientific operations depends on the effectiveness of the data flow from Kenya to SSDC and the data analysis and software management.


\bibliography{P4-24}


\end{document}